\documentclass[aps,prl,twocolumn,superscriptaddress,a4paper]{revtex4}

\usepackage{graphicx}

\pdfoutput=1 

\usepackage{fancyhdr}
\usepackage[colorlinks=true,citecolor=blue,linkcolor=magenta]{hyperref}
\hypersetup{%
pdfauthor = {Patrick Maletinsky},
colorlinks = true, linkcolor = blue, urlcolor=blue, bookmarksnumbered =  true}

\begin{document}


\title{Quantum control of proximal spins using nanoscale magnetic resonance imaging}


\author{M.S. Grinolds\footnote{These authors contributed equally to this work}}
\affiliation{Department of Physics, Harvard University, Cambridge, Massachusetts 02138 USA}
\author{P. Maletinsky\footnotemark[\value{footnote}]}
\affiliation{Department of Physics, Harvard University, Cambridge, Massachusetts 02138 USA}
\author{S. Hong\footnotemark[\value{footnote}]}
\affiliation{School of Engineering and Applied Science, Harvard University, Cambridge, Massachusetts, 02138 USA}
\author{M.D. Lukin}
\affiliation{Department of Physics, Harvard University, Cambridge, Massachusetts 02138 USA}
\author{R.L. Walsworth}
\affiliation{Harvard-Smithsonian Center for Astrophysics, Cambridge, Massachusetts 02138 USA}
\author{A. Yacoby}
\affiliation{Department of Physics, Harvard University, Cambridge, Massachusetts 02138 USA}
\email[]{yacoby@physics.harvard.edu}

\date{\today}

\begin{abstract}
Quantum control of individual spins in condensed matter systems is an emerging field with wide-ranging applications in spintronics\,\cite{Awschalom2002}, quantum computation\,\cite{Nielsen2000}, and sensitive magnetometry\,\cite{Chernobrod2005}. Recent experiments have demonstrated the ability to address and manipulate single electron spins through either optical\,\cite{Jelezko2004,Xu2007} or electrical techniques\,\cite{Hanson2007,Morello2010,Foletti2009}. However, it is a challenge to extend individual spin control to nanoscale multi-electron systems, as individual spins are often irresolvable with existing methods. Here we demonstrate that coherent individual spin control can be achieved with few-nm resolution for proximal electron spins by performing single-spin magnetic resonance imaging (MRI), which is realized via a scanning magnetic field gradient that is both strong enough to achieve nanometric spatial resolution and sufficiently stable for coherent spin manipulations. We apply this scanning field-gradient MRI technique to electronic spins in nitrogen-vacancy (NV) centers in diamond and achieve nanometric resolution in imaging, characterization, and manipulation of individual spins. For NV centers, our results in individual spin control demonstrate an improvement of nearly two orders of magnitude in spatial resolution compared to conventional optical diffraction-limited techniques. This scanning-field-gradient microscope enables a wide range of applications including materials characterization, spin entanglement, and nanoscale magnetometry.
\end{abstract}

\maketitle

Magnetic field gradients allow for spatially distinguishing spins in ensembles, as fields locally modify the spins' resonance frequencies. Spatially separated spins can therefore be addressed selectively, allowing for magnetic resonance imaging (MRI), which has revolutionized the medical and biological sciences by yielding micron-scale imaging of nuclear spins\,\cite{Mansfield2005,Lee2001}. Performing MRI on single spins with high spatial resolution is attractive both for determining structure on the molecular scale and for achieving individual spin quantum control in ensemble systems. With conventional MRI techniques, however, it is difficult to improve the spatial resolution to the nanoscale due to insufficient readout-sensitivity and inadequate magnetic field gradients\,\cite{Glover2002}. Recently, magnetic field gradients introduced via scanning probe techniques have enabled single spin detection with few-nm resolution\,\cite{Rugar2004,Balasubramanian2008}; however, control and characterization of individual spins in nanoscale clusters has not been demonstrated thus far.

Here we perform scanning field-gradient MRI on proximal electron spins in nanoscale ensembles and demonstrate a spatial resolution $<10~$nm under ambient conditions.  We show that scanning field-gradient microscopy not only allows for imaging but further provides quantum spin control for characterization and manipulation of individual spins on the nanoscale. By pushing the spatial resolution to few-nm length scales, our results illustrate that quantum control of individual spins can be maintained in dense ensembles of spins, where the mutual coupling between adjacent spins can become very strong\,\cite{Neumann2010}. Thus, scanning-field-gradient MRI will help facilitate the creation of entangled spin-states with applications for quantum information processing and sensitive, nanoscale magnetometry. While our approach is applicable to any spin system where spins can be initialized and read out, we focus here on the electronic spins associated with NV centers in diamond, where spin initialization and readout can be performed optically\,\cite{Gruber1997}. Additionally, NV spins are attractive for performing quantum information processing\,\cite{Neumann2010,Dutt2007} and sensitive magnetometry\,\cite{Balasubramanian2008,Maze2008,Taylor2008,Degen2008}. Individual NV spin control in a nanoscale ensemble is a key advance towards the implementation of these applications.

\begin{figure}
\includegraphics[scale=0.5]{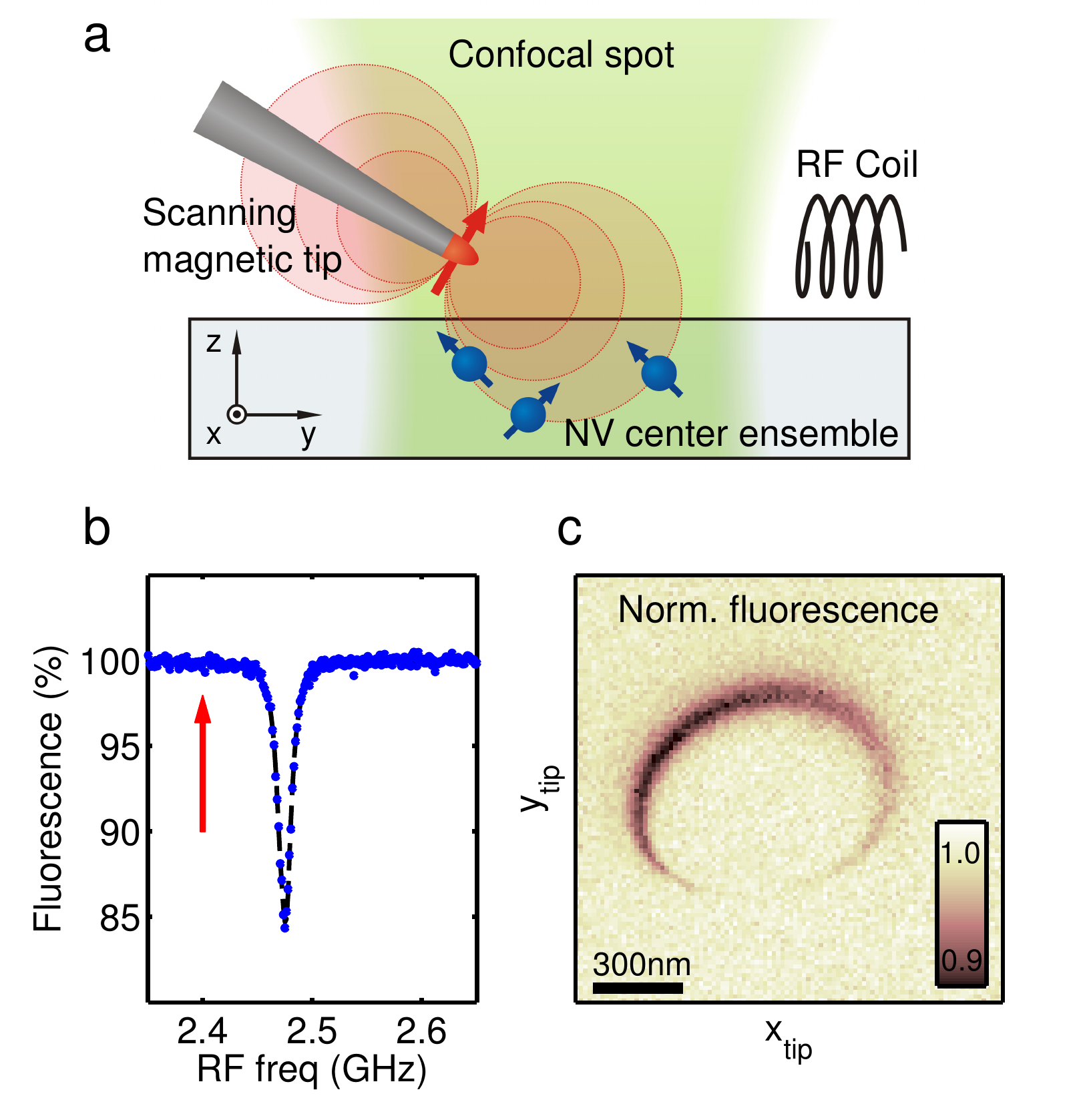}
\caption{\label{Fig1} Two-dimensional imaging of a single NV center using the scanning-field-gradient MRI microscope. (a) Schematic of the experimental approach. A magnetic tip is scanned by an AFM over a diamond sample containing multiple shallow NV centers separated by distances smaller than the optical excitation wavelength. Each NV center experiences a different magnetic field from the tip. Thus, spin transitions for individual NV centers can be selectively driven by tuning the frequency of an externally applied radio-frequency (RF) field. A confocal microscope provides NV spin-state preparation via optical pumping, and detection via spin-state dependent fluorescence. (b) Optical fluorescence measurement of the electron spin resonance (ESR) of a single NV center in the absence of the magnetic tip. The NV-spin-dependent fluorescence rate leads to a drop in emitted (and detected) photons when the external RF source is swept onto the NV spin resonance (here at 2.47 GHz instead of the zero-field value of 2.87 GHz because of an applied static magnetic field). (c) Two-dimensional magnetic resonance image of a single NV center, created by scanning the magnetic tip across the surface and fixing the RF frequency off resonance from the NV ESR transition (see arrow in (b)). A reduction of fluorescence is observed for positions of the magnetic tip relative to the NV that put the NV spin on resonance with the RF field, creating a dark ``resonance ring''. The plotted fluorescence magnitude is normalized to the NV fluorescence when no driving RF field is applied near the ESR transition.}
\end{figure}

Our scanning-field-gradient MRI system (Figure\,\ref{Fig1}a) is comprised of an atomic force microscope (AFM) outfitted with a magnetic tip and integrated into an optical confocal microscope, all operating under ambient conditions. Small ensembles of shallowly implanted NV centers, $\approx10-50~$nm below the surface of a diamond sample, are placed in the confocal spot (volume $<1~$cubic micron) where an excitation laser at $532~$nm is used to initialize and read out the NV centers' spin states. Because of the NV centers' spin-dependent fluorescence ($\approx630-800~$nm), optically detected electron spin resonance (ESR) can be observed by sweeping the frequency of a driving radio-frequency (RF) field through the spin resonance and measuring the corresponding variation in fluorescence (Figure\,\ref{Fig1}b)\,\cite{Gruber1997}, in this case on a single NV center. As has been demonstrated by \emph{Balasubramanian et al.}\,\cite{Balasubramanian2008}, a magnetic tip in proximity to an NV center shifts the energy of the NV spins, particularly for fields along the NV axis. By selectively detuning the applied RF field and scanning the magnetic tip over a single NV center, a magnetic field map corresponding to the applied detuning can be acquired (Figure\,\ref{Fig1}c). Using the AFM to regulate the magnetic tip's height allows for precise nanoscale control of the tip's location in three dimensions: the height of the tip is maintained via a conventional feedback loop on the force from the sample, while several-nm precision in lateral positioning is achieved by using the sample's topography as a marker.

\begin{figure*}
\includegraphics[scale=0.5]{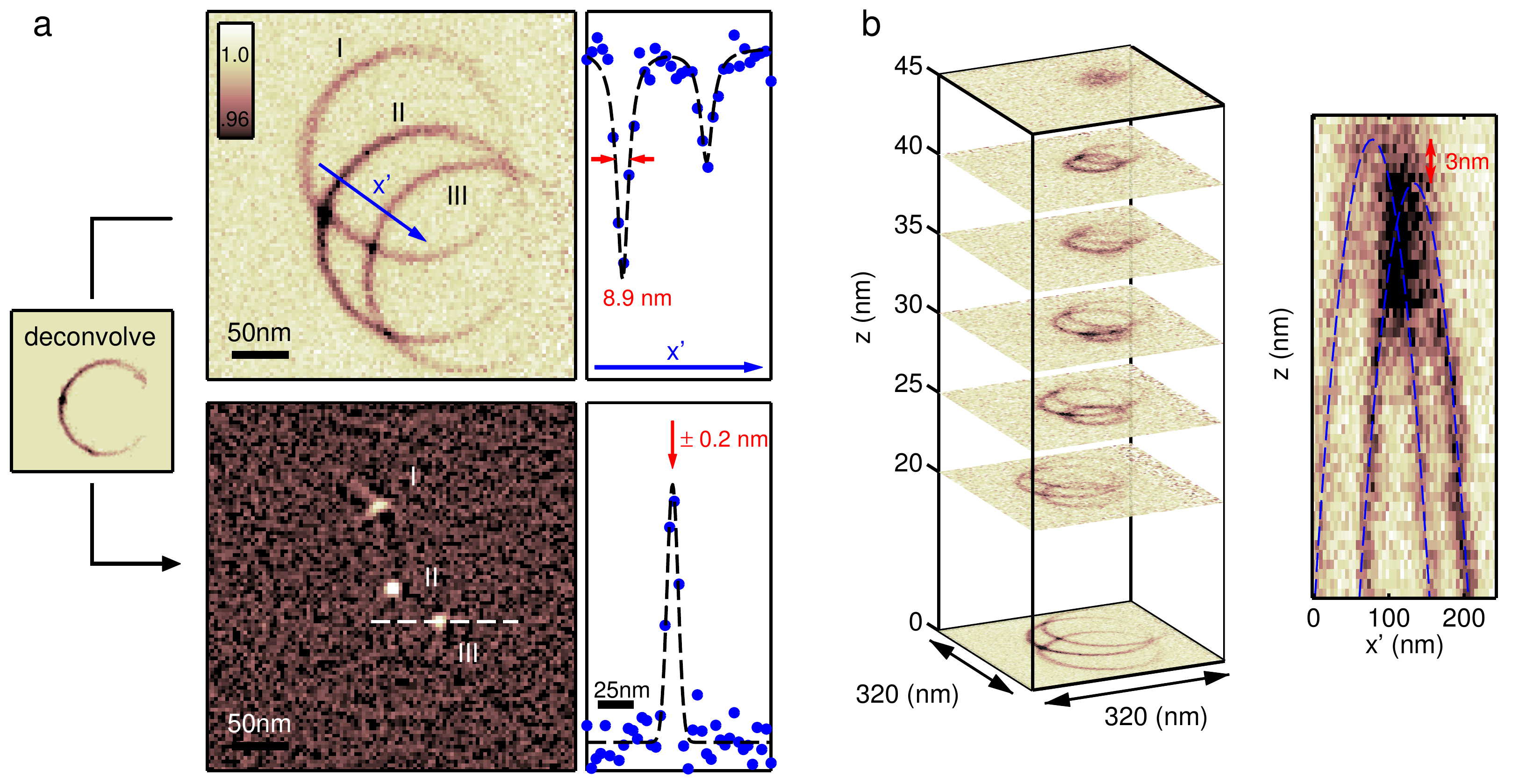}
\caption{\label{Fig2} Three-dimensional magnetic resonance imaging of proximal NV spins. (a) Scanning the magnetic tip over a cluster of three NV centers with the same crystallographic orientation yields multiple dark-resonance rings in the observed fluorescence, one for each NV center (upper image). The relative locations of the NV centers are extracted from the magnetic resonance image through a deconvolution process, yielding adjacent NV-NV distances of $50~$nm and $70~$nm (lower image). ESR spectral linecuts (e.g., along x' as shown here) give a spatial resolution of roughly $9~$nm with $0.2~$nm precision taken from a two-dimensional Gaussian fit to the deconvolved peak. (b) The relative depths of the NV centers below the diamond surface (i.e., relative NV positions along the z-axis) are determined by taking magnetic resonance images for different magnetic tip heights above the sample surface. Comparing the evolution of the three NV resonance rings as a function of tip-to-sample distance, we determine that NV I lies roughly $15~$nm below NV II and NV III (stack of images on left). A vertical cut further resolves NV II and NV III, showing that they are $3~$nm apart in depth (right image).}
\end{figure*}

We first demonstrate how this technique can be used for imaging proximal NV spins. The magnetic response shown in Figure\,\ref{Fig1}c provides a direct means for determining the relative location of proximal NV centers. For NVs with an identical orientation, the indistinguishable nature of their spins leads to an identical but spatially shifted magnetic field map for each NV spin. Thus, magnetic resonance images of single-spins (such as in Figure\,\ref{Fig1}c) can serve as the point spread function for multi-spin imaging. To demonstrate the performance of this technique, we executed scanning-field-gradient imaging on three closely spaced NV centers (Figure\,\ref{Fig2}). As the magnetic tip is scanned laterally across the sample, a magnetic field contour is observed for each center. Here, we selected an ensemble of NVs where all three have a common quantization axis orientation, so that every spin responds in the same way to the presence of the magnetic tip. Thus, the relative distances between NV spins can be obtained by quantifying the spatial shift between the observed resonance rings through a deconvolution procedure. The resulting image (Figure\,\ref{Fig2}a) indicates the relative positions of the NV centers, which we find are spaced by $50~$nm and $75~$nm with respect to NV II. A spatial resolution of $9~$nm can be extracted from the width of the resonance ring along the vector connecting NV centers II and III; a precision of $0.2~$nm can be determined from a two-dimensional Gaussian fit to the deconvolved peak. This estimate of precision only accounts for variance induced by random noise and does not account for systematic deviations caused by effects such as scanner non-linearity or deviations from the peak shape to the Gaussian fit.

Our single-spin MRI technique can be extended to three dimensions by performing magnetic tomography, where the magnetic tip is retracted from the surface in few-nm steps and scanned laterally (Figure\,\ref{Fig2}b). The size of the resonance rings evolves quickly as a function of z-distance as the magnetic field from the tip becomes too weak to bring the spin-transitions into magnetic resonance with the detuned RF driving field. The height differences between the NVs can be seen here as NV I vanishes roughly $15~$nm before the other two, which indicates that NV I lies roughly $15~$nm further below the surface than both NV II and III. A vertical scan across NVs II and II shows the z-resolution of this measurement to be $10~$nm, which is extracted from the width of the resonance line in the z-direction. Since the lateral separation of these two NVs is larger than our spatial resolution, we can determine the relative height of the two NVs with high precision, which we find to be $3~$nm.

The NV centers in our demonstration experiments were created through implantation of nitrogen ions forming a layer $\approx10~$nm below the diamond surface. Modeling of this implantation procedure\,\cite{Ziegler2010} predicts a spatial variation of $\pm3~$nm; whereas we observe a larger variation in the distribution of NV depths ($>10~$nm) using our scanning-field-gradient MRI technique. This discrepancy may arise from ion channeling or surface effects. Nanoscale precision in measuring the distance between proximal NV spins also allows determination of the mutual dipole coupling between adjacent spins, a key component for creating entangled spin-states.

Scanning-field-gradient MRI provides not only a method for nanoscale mapping of the spatial locations of proximal spins but also allows for individual spin transitions to be resolved in frequency space with high precision. In the absence of magnetic field gradients, identical spins sharing a quantization axis are indistinguishable making selective control over proximal spins impossible. Performing scanning-field-gradient MRI on proximal spins differentiates their transition frequencies, allowing for coherent manipulation and characterization of individual spins in the ensemble. This can be done simultaneously on several NV spins while preserving independent control of each spin. Such selective control over proximal spins can be maintained as long as the magnetic field gradient separates the transition frequencies of neighboring spins by more than their resonance linewidth, which is the same condition that determines the spatial resolution in MRI. Therefore, spins separated by more than the achieved MRI spatial resolution ($\approx9~$nm for the NV spin experiments reported here) can be addressed independently using our technique.

\begin{figure}
\includegraphics[scale=0.5]{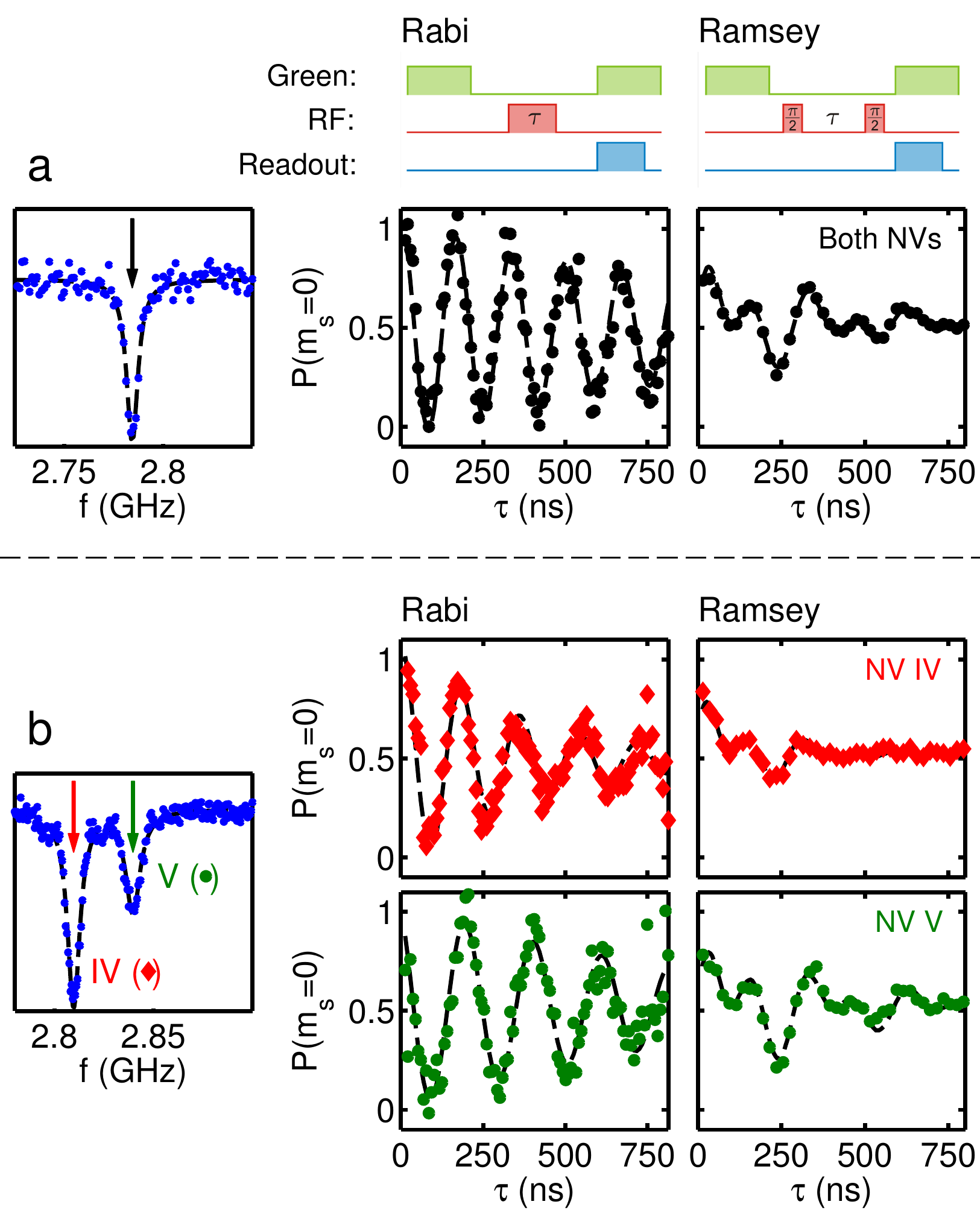}
\caption{\label{Fig3} Individual quantum control and characterization of proximal NV centers. (a) With the magnetic tip pulled far away from the sample, two NV centers (IV and V) separated by $135~$nm and sharing the same spin-quantization axis cannot be resolved by their ESR spectra ($m_S=0 \rightarrow m_S=1$ transition of the NV triplet groundstate, observed by spin-state dependent fluorescence), as they experience the same static magnetic field (left panel). By driving both NVs at once, Rabi oscillations and the NV spins' collective free-induction decay can be observed using the indicated Rabi and Ramsey RF pulse sequences, respectively; but the evolution of each individual NV-spin remains undetermined. (b) With the magnetic tip in close approach to the sample, the two NV spin resonances are spectrally distinguishable by their differing Zeeman shifts ($\approx30~$MHz splitting), allowing each NV to be addressed independently, so that Rabi oscillations and Ramsey free-induction decay of each NV spin can be individually measured (NV IV in red and NV V in green). For these measurements, the probability of the spin being in the $m_S=0$ state is plotted. We observe that NV IV has a shorter inhomogeneous coherence time ($T_2^*\approx180~$ns) than NV V ($T_2^*\approx420~$ns). The Ramsey free induction decay measurements are taken with the RF driving field detuned by $5~$MHz from the target NV center's nominal ESR frequency in the presence of the magnetic tip; the observed oscillations are due to beating of this detuning with the 15N hyperfine splitting ($3.1~$MHz). The measured Ramsey data are fit to the sum of two exponentially damped sinusoids, whose phases are fixed by the relative strength between the net detuning (the RF field detuning plus or minus half the hyperfine splitting) and the Rabi frequency ($5.5~$MHz), governed by the strength of the applied RF field.}
\end{figure}

To demonstrate such selective nanoscale characterization and control of proximal spins we examined a pair of proximal NV centers (NV IV and NV V, separated by $135~$nm, which share the same NV axis orientation. In the absence of the magnetic tip, we performed a continuous-wave ESR measurement on both NV spins simultaneously (Figure\,\ref{Fig3}a, left panel), finding no difference in their spectra, as expected. After tuning the RF frequency to the pair of NV spins, we drove Rabi oscillations and observed the free induction decay of the two NV spins via a Ramsey sequence (Fig\,\ref{Fig3}a, right panels). The measured Ramsey fringes show a pronounced beating pattern due to the hyperfine structure of the NV spin transitions and are damped due to inhomogeneous dephasing of the two NV centers. However, information distinguishing the individual coherence-properties of the proximal NV centers cannot be extracted from these measurements. In contrast, when the magnetic tip is brought in close proximity to the NV spins, it splits their resonance frequencies due to the differing magnetic fields applied to each NV, thereby allowing each NV spin to be addressed and characterized individually and coherently (Figure\,\ref{Fig3}b) without modifying the spin of the neighboring NV.

Such coherent individual spin control was realized by tuning the RF frequency to the ESR resonance of the target NV center. Pulsing the RF with a variable duration induced coherent Rabi oscillations of either NV IV or NV V, dependent on the tuning of the RF frequency. Similarly, characterization of individual spin coherence properties was achieved by measuring the spin's free-induction decay via a Ramsey sequence using the appropriate RF frequency for the target NV spin. We observed that NV IV has a faster free-induction decay rate than NV V, indicating that the two spins have different inhomogeneous dephasing times, $T_2^*$. The measured free-induction decay rate for NV V with and without the magnetic tip are comparable, indicating that additional decoherence induced by the tip is small compared to that spin's ambient dephasing $1/T_2^*$. In order to prevent tip-induced spin-decoherence, we employed a three-dimensional spatial feedback scheme, which ensured that variations in the applied tip-field were smaller than the intrinsic NV inhomogeneous dephasing rates. Thus, the applied magnetic field gradient can be used to characterize individual spin coherence properties, a direct consequence of achieving single-spin control.

\begin{figure}
\includegraphics[scale=0.5]{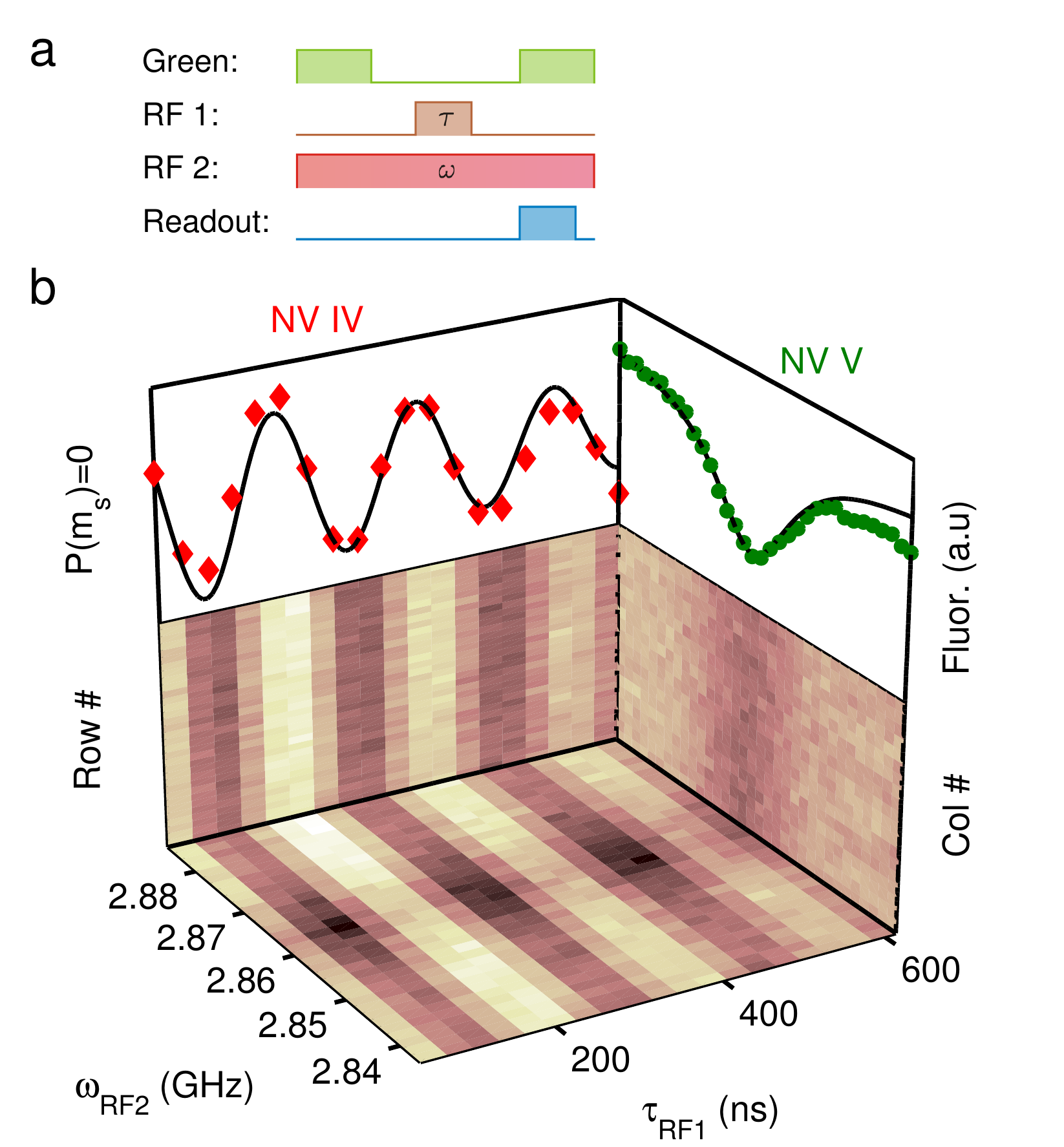}
\caption{\label{Fig4} Selective, independent RF control of proximal NV spins in the presence of the tip magnetic field gradient. (a) Fluorescence measurements are performed for the same two proximal NV centers as in Figure\,\ref{Fig3}, while undergoing simultaneous, near-resonant driving by fields RF 1 and RF 2; and with the tip magnetic field gradient inducing a large ($\approx30~$MHz) Zeeman frequency shift between the ESR frequencies of the two NV spins. The frequency of RF 1 is set on resonance with NV IV and is pulsed with varying duration ($\tau_{RF1}$) to induce Rabi oscillations. Simultaneously, RF 2 is continuously applied while its frequency ($\omega_{RF2}$) is swept through the spin resonance of NV V to measure its ESR spectrum. (b) To determine if these simultaneous NV measurements are independent, a two-dimensional data set containing all relevant values of both $\tau_{RF1}$ and $\omega_{RF2}$ is acquired (base of cube). From this data set, the individual behavior of each NV can be extracted by subtracting the mean measured fluorescence for each row (i.e., fixed value of $\omega_{RF2}$) from the data for that row; and similarly for each column (fixed value of $\tau_{RF1}$). The resulting extracted data sets (shown in the sides of the cube) are found to be independent of row or column number, showing that varying the Rabi pulse duration $\tau_{RF1}$ for NV IV does not influence the results of sweeping $\omega_{RF2}$ through the spin resonance of NV V, and vice versa. The only observed deviations from this independent NV spin control are for small values of $\omega_{RF2}$ where the frequency of RF 2 approaches the spin resonance of NV IV, thereby causing modest damping of this NV's Rabi oscillations. These deviations disappear near the resonance of NV V, showing that both NVs can be driven independently. Summing up the measurements for all rows and columns yields the resulting averaged Rabi oscillation measurement for NV IV (top of left wall) and ESR spectrum from NV V (top of right wall), respectively.}
\end{figure}

Under the influence of the tip's magnetic field gradient, manipulating one NV spin does not perturb the state of a neighboring NV spin. To experimentally verify this selective, independent spin control, we simultaneously drove Rabi oscillations on NV IV while measuring the ESR spectrum of NV V (Figure\,\ref{Fig4}a) by applying two separate RF fields (RF 1 and RF 2), each used to drive one NV spin-resonance. To illustrate the independence of the two measurements, we performed a two-dimensional sweep over the RF 1 pulse duration ($\tau_{RF1}$) and the frequency of RF 2 ($\omega_{RF2}$) (Figure\,\ref{Fig4}b). We then subtracted out the mean of each row or column to show that the two measurements are independent. Rows or columns were summed to reconstruct the resulting Rabi oscillations of NV IV or the ESR curve of NV V, respectively. The only deviations from simultaneous, independent NV spin measurements are observed when $\omega_{RF2}$ is swept close to the resonance of NV IV, yielding a slight damping in the Rabi oscillations. These deviations do not persist through the ESR transition of NV V, which shows that both NVs can be manipulated independently when they are driven on resonance.

Scanning-field-gradient MRI of spins yields precise determination of their relative locations with nanometric spatial resolution, which will be crucial for optimizing the performance of spin-based magnetometers and the functionality of spin-based quantum bit ensembles. Once suitable spin ensembles are identified and spatially mapped, individual spin control both allows for the determination of individual spin properties and - when combined with magnetic dipole coupling between adjacent spins - provides a method for achieving complete control of the quantum state of spin ensembles. The spatial resolution of our scanning technique is extendable to the atomic scale by using stronger magnetic field gradients\,\cite{Mamin2007} and narrower ESR linewidths\,\cite{Balasubramanian2009}. For NV spins, the current experimental barrier to improving this spatial resolution is overcoming the reduction in ESR contrast due to a strong off-axis magnetic field created by the tip\,\cite{Epstein2005}; however, this can be prevented by either applying a strong bias-field along the NV axis or by tailoring the domain structure of the magnetic tip to produce high gradients but only moderate total fields.

The control and manipulation of individual spins using magnetic field gradients is independent of the method used for spin readout. For optically addressable spins, such as NV spins, integrating far-field, sub-diffraction schemes - such as stimulated emission depletion (STED)\,\cite{Rittweger2009} and reversible saturable optical linear fluorescence (spin-RESOLFT)\,\cite{Maurer2010} - with a scanning magnetic field gradient would allow for both robust individual spin control and readout with nanometric resolutions. Additionally, selective optical control of such systems is possible via the incorporation of an electric field gradient to the scanning tip\,\cite{Hettich2002}, which would allow both spin and electronic degrees of freedom to be both addressed individually. Alternatively, using demonstrated single-shot electrical readout of individual spins\,\cite{Morello2010,Barthel2009} would allow for MRI to be performed rapidly and efficiently, as acquisition times would not be limited by the readout integration time. Individual control of spins via the present technique is also extendable to nuclear spins, provided methods for reading out nuclear spins reach single-spin sensitivity\,\cite{Mamin2007}. Nanoscale or atomic spatial resolution of nuclear spins is feasible as their long spin coherence times help to compensate for nuclear spins' small dipole moment. For any spin system, providing individual control of spins in dense ensembles, where mutual coupling is strong, allows for the creation of arbitrary entangled states. Such states have intriguing potential applications ranging from sensitive nanoscale magnetometers to scalable quantum information processors\,\cite{Yao2010}.

We gratefully acknowledge G. Balasubramanian and P. Hemmer for fruitful technical discussions, as well as B. Hausmann and M. Loncar for instruction in the fabrication of NV center containing nanostructures. M.S.G. is supported through fellowships from the Department of Defense (NDSEG program) and the NSF. P.M. acknowledges support from the Swiss National Science Foundation, and S.H. thanks the Kwanjeong Scholarship Foundation for fellowship funding. This work was supported by NIST and DARPA.


Correspondence and requests for materials should be addressed to A.Y. (\href{mailto:yacoby@physics.harvard.edu}{yacoby@physics.harvard.edu})
\\
\\
\\
\\

\bibliographystyle{apsrev4-1}
\bibliography{BibQuantContr}

%
\section*{Methods}

\subsection*{NV center samples:}
NV centers were created through implantation of 15N ions\,\cite{Rabeau2006} into ultrapure diamond (Element Six, electronic grade diamond, $<5~$ppb nitrogen). The implantation was done at $6~$keV to give a nominal nitrogen depth of $10~$nm\,\cite{Ziegler2010}. To form NV centers, the sample was annealed in vacuum at $750~^\circ$C, where existing vacancies are mobile and can pair with the implanted nitrogen atoms. The resulting density of NV centers corresponds to one center every 50-100 nanometers, forming a layer roughly $10~$nm from the surface. To isolate small NV clusters (i.e., a few proximal NV centers), we selectively etched\,\cite{Lee2008} away the majority of the shallow diamond surface layer, leaving proximal NV-containing nanostructures. This was done using electron beam lithography to define an etch mask from a flowable oxide (Dow Corning, XR-1541)\,\cite{Hausmann2010}. A reactive-ion etch then removed any exposed diamond surfaces, resulting in shallow diamond nanostructures ($100-800~$nm across) containing ensembles of proximal NV spins.
\subsection*{Magnetic tips:}
Magnetic tips were created by evaporating a magnetic layer onto quartz tips of roughly 80 nm in diameter, which were fabricated using a commercial laser-pulling system (Sutter Instrument Co., P-2000). Using a thermal evaporator, a $25~$nm layer of cobalt-iron was deposited on the side of the pulled quartz tip. A $5~$nm chrome layer was then evaporated, which serves as a capping layer to prevent oxidation of the magnetic material. These tips result in magnetic field gradients of roughly 1G/nm at distances of roughly $100~$nm.
\subsection*{Magnetic tip positioning with an AFM:}
To achieve high spatial resolution in NV imaging and manipulation, it is necessary to control precisely the relative distance between the magnetic tip and the addressed NV centers. Such nanoscale control can be challenging under ambient conditions (standard temperature and pressure); e.g. temperature drifts of a small fraction of a degree can induce few nm drifts between the magnetic tip and the diamond sample, which would inhibit local spin-control.

To overcome this problem, we used an AFM to position the magnetic tip in three dimensions with precisions of a fraction of a nanometer in z and a few nanometers in both x and y. Height control was achieved through normal AFM operation in which a feedback loop modulates the height of the tip to keep the sample-tip interaction constant. Lateral positioning was achieved through intermittently locating topographic features on the sample to ensure that the tip's relative distance to the sample is fixed during any performed measurements.

\end{document}